\documentclass[sigconf]{acmart}
\usepackage{enumitem}
\usepackage{booktabs}
\usepackage{colortbl}

\definecolor{cellgreen}{HTML}{C8EFC0}
\definecolor{cellpink}{HTML}{FFC7B8}

\usepackage{array}
\newcolumntype{P}[1]{>{\raggedright\arraybackslash\hyphenpenalty=10000\exhyphenpenalty=10000}p{#1}}

\AtBeginDocument{%
  }

\setcopyright{acmlicensed}

\copyrightyear{2026}
\acmYear{2026}
\setcopyright{cc}
\setcctype{by}
\acmConference[CHIWORK '26]{Proceedings of the 5th Annual Symposium on Human-Computer Interaction for Work}{June 22--25, 2026}{Linz, Austria}
\acmBooktitle{Proceedings of the 5th Annual Symposium on Human-Computer Interaction for Work (CHIWORK '26), June 22--25, 2026, Linz, Austria}
\acmDOI{10.1145/3808045.3808066}
\acmISBN{979-8-4007-2429-9/2026/06}

\begin{document}

\title{From 911 to Hospital: Challenges and Opportunities for AI Integration in Emergency Medical Services}

\author{Emily Hou}
\affiliation{%
  \institution{Wellesley College}
  \country{USA}
}

\author{Marelyn Gonzalez}
\affiliation{%
  \institution{Wellesley College}
  \country{USA}
}

\author{Andrew L. Kun}
\affiliation{%
  \institution{University of New Hampshire}
  \country{USA}
}

\author{Osnat Mokryn}
\affiliation{%
  \institution{University of Haifa}
  \country{Israel}
}

\author{Orit Shaer}
\affiliation{%
  \institution{Wellesley College}
  \country{USA}
}



\begin{abstract}
Artificial Intelligence (AI) is increasingly introduced into healthcare settings, yet its integration into fast-paced, high-pressure domains such as Emergency Medical Services (EMS) remains limited. EMS work unfolds across distinct stages, each characterized by different information needs, constraints, and forms of collaboration. Designing effective AI support requires understanding how AI interventions align with, or disrupt, EMS work across its different stages.

We conducted semi-structured interviews with 25 EMS clinicians across the United States to examine how existing technologies currently support emergency services workflows and how they envision opportunities for, and concerns about, future AI-based support across different stages of emergency response. Our analysis reveals the cognitive, social, and procedural factors that enable EMS team coordination, which is grounded in \textit{situational awareness} across distributed roles. EMS clinicians expressed significant concerns about how AI integration threatens this coordination mechanism across multiple dimensions: legal and privacy issues, technical reliability, contextual sensitivity, professional autonomy, and workflow friction. We propose five design principles for AI systems that augment distributed cognition and situational awareness, enabling EMS teams to deliver effective care under extreme constraints.


\end{abstract}




\maketitle

\section{Introduction}

Emergency medical services (EMS) clinicians provide time-critical pre-hospital care to patients 
\cite{jensen2011paramedic, nasem2012_crisis_ems}. The distinguishing characteristic of their work is the context in which it is performed: they provide medical care outside of the ordered setting of a hospital, often in uncontrollable, chaotic, and even dangerous circumstances \cite{campeau2008space}.

Many aspects of the work of EMS clinicians have been studied. This includes understanding the stages of the emergency response, from receiving a call for help, to arriving at the scene, to on-scene treatment of patients, to transporting patients to a medical facility \cite{jensen2011paramedic}. It also includes clinical decision making in the field, under time pressure, when patients' well-being is at stake \cite{jensen2011paramedic, perona2019paramedic}. Furthermore, researchers have explored the use of technology by EMS clinicians, including communications technology \cite{zhang2020evaluative, janerka2023prehospital}, diagnostic technology \cite{campbell2005prehospital,lumley2020scoping, hartline2025army}, as well as technology to support dispatch operations \cite{shekhar2025use}.

However, the exploration of technology is often focused on (important) constrained problems, applications, and techniques. Yet, we believe there is an opportunity to provide a systematic description of technology use by EMS clinicians that can tell us which technologies are used in which phases of the emergency response, and how the technology use taxes the cognitive resources of the EMS clinicians. Furthermore, we see a parallel to the early 2000s, when advances in computing hardware made it possible to deploy computers in first responder vehicles in the field \cite{kun2015user, kun2005computers}. Similarly, today's advances in AI might make it possible to deploy advanced software in support of EMS clinicians in field. We see an opportunity to use a systematic description of technology use in exploring ways how AI can be useful for EMS clinicians.

AI has rapidly transformed healthcare decision-making
~\cite{rajpurkar2022ai}, yet its integration into EMS remains limited and fraught with sociotechnical complexity. 
Research in the Human–Computer Interaction (HCI), Computer-Supported Cooperative Work (CSCW), and CHIWORK communities has increasingly focused on how AI transforms the nature of work. This growing body of work highlights that the question is not only whether AI can perform tasks, but how it can integrate into teams, workflows, and decision-making under pressure~\cite{zhang2024rethinking, wen2025trust, chen2025engaging}. Within CHIWORK, such investigations are essential to understanding how automation and AI reshape accountability, coordination, and expertise in high-stakes domains \cite{Solovey_CHIWork25}.

Emergency Medical Services represent a critical, underexplored site for such inquiry. 
Previous work has shown that paramedics rely on contextual cues and heuristic reasoning under pressure, shifting from analytical to intuitive modes of thought during emergencies~\cite{perona2019paramedic}. However, little is known about how these cognitive and organizational factors interact with emerging AI systems, or how EMS clinicians perceive the opportunities and threats of AI integration.

\subsection{Defining AI in the EMS context}
Throughout this paper, we use \textit{AI} to refer to a family of computational systems that learn patterns from data to perceive, interpret, generate, or recommend in ways that previously required human judgment. Within the scope of EMS workflows, four categories are most relevant: (1) large language models (LLMs) and conversational systems built on them, including retrieval-augmented systems that ground outputs in clinical literature (e.g., Open Evidence) and LLM-based triage and dispatch tools \cite{masanneck2024triage, shekhar2025use, shen2025enhancing}; (2) predictive machine learning models that estimate risk, classify physiological states, or interpret structured clinical data such as vital signs trajectories or EKG waveforms; (3) computer vision systems that analyze images or video, including diagnostic imaging support \cite{hartline2025army, yu2024heterogeneity} and assistance with scene or specimen capture; and (4) speech and audio processing systems, including automatic speech recognition for transcription and interpretation of 911 caller speech. 

We distinguish AI from automation, which we use to refer to deterministic systems that execute fixed procedures without learned components, such as an automatic CPR device, a powered loading stretcher, or a calculator. However, the boundary is not always sharp: some systems that EMS clinicians described to us combine learned and rule-based components, and the rule-based clinical decision support tools of earlier generations have historically been grouped under AI. 

\subsection{Research Questions}
To address the gaps identified above, we conducted semi-structured interviews with 25 EMS clinicians across the United States, representing municipal, private, and volunteer organizations. The US EMS system, in which non-physician clinicians deliver protocol-driven care, presents a particularly compelling site for examining how AI integration intersects with distributed coordination and professional expertise. This study explores three key research questions:

\begin{enumerate}
[label=\textbf{RQ\arabic*}]
\item How do EMS clinicians currently use technology as they operate in high-stress, high-stakes situations, and what cognitive resources are involved?
\item What challenges do EMS clinicians face across different stages of emergency response workflows?
\item What do EMS clinicians perceive as the risks and opportunities for integrating AI-powered technology support into EMS workflows?

\end{enumerate}

Our analysis reveals how the EMS workflow is shaped by contextual constraints, team communication patterns, and professional norms. We further identify how EMS clinicians imagine and evaluate the introduction of AI decision support, emphasizing concerns around trust, autonomy, and situational awareness alongside perceived benefits such as improved triage accuracy, coordination, and training.

This paper makes three contributions to CHIWORK and the broader HCI community:

\begin{itemize}
    \item An empirical characterization of EMS operation and current use of technology under uncertainty and time pressure, illuminating cognitive heuristics, team dynamics, and workflow constraints.
    \item An analysis of frontline perceptions of AI-support, outlining perceived risks and benefits across organizational contexts and EMS workflows.
    \item Design implications for human-centered AI systems that improve workflows and alleviate challenges in emergency care.
\end{itemize}

By situating these findings within CHIWORK’s sociotechnical lens, we extend understanding of how AI reshapes the future of work in high-reliability, safety-critical settings where human expertise and collective coordination remain indispensable.

\section{Related Work}
\subsection {Clinical AI-Powered Decision Support Systems: Sociotechnical Approach}
AI-powered healthcare decision-making systems have demonstrated the potential to improve efficiency and accuracy across diverse medical settings~\cite{rajpurkar2022ai}. Yet, incorporating them is challenging. The complexity of healthcare workflows requires technologies that not only address technical challenges but also fit seamlessly within the professional, collaborative, and often high-stress contexts in which they are deployed. For example, patient care information systems in hospitals are designed to streamline workflows by centralizing and organizing patient data, yet their success depends on seamless integration into established routines and practices~\cite{berg1999patient}. Systems also need to account for cultural norms and adapt to dynamic conditions of high-stakes environments~\cite{mebrahtu2021effects} and to changing conditions such as low-resource settings ~\cite{yu2024heterogeneity}.  

In emergency medical situations, decision-making is often shaped by uncertainty and time constraints. Tools that are perceived as cumbersome or time-consuming, or that fail to inspire trust are often disregarded~\cite{harenvcarova2017managing,mebrahtu2021effects,jacobs2021designing}.
Healthcare professionals are more likely to adopt systems that provide clear, interpretable outputs that can be cross-referenced with their own expertise~\cite{beede2020human}.

\subsection{Making Decisions Under Stress } Time pressure and complexity exacerbate cognitive strain, leading decision-makers to adopt simplifying strategies, often resulting in suboptimal decisions. Stress diverts cognitive resources, which may lead to rejection of correct solutions, delayed decisions, or an overreliance on heuristics~\cite{phillips2020decision}. Stress leads to reduced focus on situational context and lower cognitive flexibility, emphasizing the need for decision support tools to reduce ambiguity and support clear, actionable insights~\cite{loftus2020artificial}. 

In this paper, we focus specifically on acute operational stress — the short-duration, high-intensity stress response triggered by active emergency response rather than the chronic occupational stress or cumulative traumatic stress that also affect EMS clinicians over their careers \cite{regehr2017ptsd}. Whether such acute stress facilitates or impairs performance depends on the clinician's cognitive appraisal of the demands and their available resources: situations appraised as challenges (resources sufficient) can mobilize attention and motor performance, while those appraised as threats (resources insufficient) tend to degrade working memory, recall, and global task organization \cite{LeBlanc_Regehr_Tavares_Scott_MacDonald_King_2012}. In a simulated cardiac scenario with paramedics, LeBlanc et al. \cite{LeBlanc_Regehr_Tavares_Scott_MacDonald_King_2012} found that, while well-rehearsed checklist actions were preserved under high-stress conditions, global performance ratings, communication quality, and case recall all declined — a pattern consistent with our participants' reports of "falling to the level of training" and with the tunnel-vision phenomenon we describe in Section 4.2.2.

Human-AI collaboration under time pressure creates additional constraints. Studies found that limited observation time before interacting with AI increases user reliance on AI suggestions~\cite{swaroop2024accuracy}, while sufficient decision time after receiving AI input allows for better evaluation and lowers blind adherence to AI recommendations~\cite{jacobs2021designing,cao2023time}. Further, different cognitive demands of tasks affect the way people interact with AI, raising the need to design systems tailored to the unique requirements of emergency medical decision-making~\cite{cao2023time,swaroop2024accuracy}.

\subsection{Emergency Medical Services as Distributed, Technology-Mediated Work
}
Emergency Medical Services (EMS) is a prototypical high-stakes, distributed work domain in which teams must coordinate rapidly across organizational boundaries under extreme uncertainty and time pressure. From initial 911 dispatch through patient handoff at receiving hospitals, emergency response involves complex information flow across heterogeneous actors: dispatch operators, paramedics, firefighters, police, and hospital staff. Each of these groups has different information needs, technological systems, and decision-making authority~\cite{hutchins1995,schmidt_bannon1992}. Understanding how these distributed teams maintain coordination through current technologies is essential for anticipating how AI integration might enhance or undermine their effectiveness.

The EMS workflow unfolds across distinct stages, each requiring different forms of coordination~\cite{jensen2011paramedic}. \textit{Dispatch operators} use Computer-Aided Dispatch (CAD) systems to gather caller information and transmit preliminary details to responding units via Mobile Data Terminals (MDTs) and radio. \textit{En route}, paramedics receive updates while building initial mental models of the scene. \textit{Upon arrival}, responders rapidly assess hazards, patient conditions, and resource needs while coordinating with fire and police through radio systems. The \textit{on-scene treatment} phase exhibits the highest coordination demands as teams integrate information from multiple sources. These include visual assessment, patient history, vital signs monitors, and protocol applications. At the same time, they manage care, equipment, and communication with medical direction~\cite{jensen2011paramedic,perona2019paramedic}. Finally, \textit{patient handoff} at hospitals requires transmitting accumulated information across organizational boundaries through verbal reports and electronic Patient Care Reports (ePCRs).  

This workflow reveals a fundamental coordination challenge: How do distributed teams working across organizational boundaries, with heterogeneous information systems and under severe time pressure, maintain sufficient shared understanding to coordinate effectively? 
Current EMS technologies (CAD systems, MDTs, radio communications, ePCRs, protocol apps, vital signs monitors) mediate how information is perceived, communicated, and understood across the system. Yet each introduces potential coordination failures: radio channel congestion, MDT display limitations in dynamic scenes, incompatible health record systems during handoffs, and protocol applications unable to accommodate atypical presentations~\cite{wilson2023scoping}. An additional question arises: How will AI systems impact the mechanisms through which distributed teams currently maintain the shared understanding necessary for effective coordination?

EMS systems vary significantly across countries in staffing models, organizational structure, and regulatory frameworks, all of which shape coordination practices and technology use. The workflow and technology analysis in this paper are situated within the US EMS context, where care is delivered by EMTs and paramedics operating across municipal, private, and volunteer agencies; we discuss plans to analyze non-US EMS contexts 
in Section 6.

\subsection{Distributed Cognition: Intelligence Across People, Artifacts, and Environments
}
Hutchins' foundational work on distributed cognition provides essential grounding for understanding coordination in complex sociotechnical systems~\cite{hutchins1995}. Through an ethnographic study of Navy ship navigation, Hutchins demonstrated that cognitive processes extend beyond individual minds to encompass interactions between people, artifacts, and environments. 


Recent applications of distributed cognition to acute care reveal the importance of understanding these information trajectories.  Wilson et al.~\cite{wilson2023scoping} show how emergency department operations depend on complex interdependencies among physical artifacts, digital systems, and social coordination practices. However, technology designers often fail to anticipate actual information trajectories, leading to coordination breakdowns when new systems disrupt established communication patterns. While distributed cognition describes the architecture of intelligence across people and artifacts, it does not fully account for the dynamic, real-time awareness that enables distributed actors to coordinate effectively. 

\subsection{Situational Awareness: The Fundamental Coordination Mechanism}
Situational Awareness (SA) describes the ongoing awareness that allows distributed actors to coordinate activities, maintain shared understanding, and project future states to enable proactive rather than reactive responses.

Endsley~\cite{endsley1995} defines SA as ``the perception of elements in the environment within a volume of time and space, the comprehension of their meaning, and the projection of their status in the near future’’ (p. 36). This three-level framework reveals how SA enables coordination. 

\textit{Level 1 SA, Perception,} involves detecting and attending to relevant cues, data, and information sources, the raw informational substrate that enables coordination. 
In EMS, this level encompasses perceiving patient vital signs, environmental hazards, bystander information, radio communications from dispatch and other units, and equipment status. Team members must perceive relevant cues to share them with others. Failures at Level 1 manifest as missed information: not hearing radio transmissions during high workload, not noticing deteriorating vitals while focused on procedures, not detecting environmental hazards due to attentional narrowing under stress~\cite{schulz2013}. These perceptual failures cascade through distributed systems. When team members operate on different informational foundations, coordination becomes impossible.

\textit{Level 2 SA, Comprehension,} involves integrating perceived elements to understand their significance relative to operational goals
, recognizing that a constellation of vital signs indicates impending cardiac arrest, understanding that bystander behavior suggests scene safety threats, or comprehending that radio traffic indicates resource unavailability. Level 2 SA enables compatible interpretations. Comprehension failures manifest as misunderstandings. 
In distributed teams, comprehension failures are particularly dangerous because members may perceive the same cues yet comprehend different situations, creating hidden coordination failures where incompatible understandings remain undetected until actions misalign~\cite{endsley_jones2001}. 

\textit{Level 3 SA, Projection,} involves anticipating future states based on current understanding, enabling proactive coordination, the most sophisticated level that distinguishes experts from novices~\cite{endsley2015}. In EMS, Level 3 SA enables paramedics to anticipate patient deterioration before clinical indicators become obvious, to project that current treatment approaches will prove insufficient, or to foresee that scene conditions will complicate transport. 
For coordination, Level 3 SA allows team members to anticipate each other's future needs. 
Projection failures manifest as reactive rather than proactive coordination. Teams respond to problems only after they occur rather than preventing them, creating cascading failures as reactive responses consume cognitive resources needed for ongoing SA maintenance~\cite{feller2023situational}. 

Research on paramedic decision-making reveals that SA quality fundamentally determines decision quality~\cite{jensen2011paramedic,perona2019paramedic}.  Experts demonstrate superior decisions not because they use different processes but because they develop richer, more accurate SA more rapidly than novices. Experience builds pattern libraries enabling faster situation recognition, but this recognition still depends on perceiving relevant cues and comprehending their significance. Novices struggle not from lacking Recognition-Primed Decision Making capabilities~\cite{klein2017sources} but from lacking the SA foundation, both from limited pattern libraries and difficulty maintaining SA under attentional and cognitive load of emergency situations~\cite{ryan_halliwell2012}.

\subsubsection{Team SA and Shared SA: Coordination Across Multiple Levels}
In collaborative emergency work, SA operates at multiple interconnected levels~\cite{endsley_jones2001}. \textit{Team SA} refers to ``the degree to which every team member has the SA required for his or her responsibilities,''  recognizing that different roles require different SA elements. 
Team SA is achieved when each member maintains sufficient awareness for their specific role, even though SA content differs across roles.

\textit{Shared SA} represents ``the degree to which team members possess the same SA on shared SA requirements'', the subset of situational elements that multiple members need for coordination~\cite{endsley_jones2001}. 
Shared SA enables implicit coordination, as team members adjust actions based on shared understanding without requiring explicit communication.

\textit{Shared mental models} provide the foundation for shared SA~\cite{cannonbowers1993}. These are consistent internal representations of team member roles, operational procedures, equipment capabilities, and prototypical scenarios, developed through training, experience, and repeated collaboration. Shared mental models function as templates enabling team members to rapidly develop compatible SA in novel situations by recognizing situation types and activating appropriate coordinated response patterns. When teams lack shared mental models, achieving shared SA requires extensive explicit communication to align understanding, communication that may be impossible under time pressure or channel constraints. 

Current EMS technologies profoundly shape SA development and maintenance. Effective technology design for SA must address all three levels~\cite{endsley2017}.

\subsubsection{The effect of AI  on SA}
 
As AI technologies advance, there is growing interest in introducing AI-based decision support into EMS workflows. Recent systems have explored AI for enhanced dispatch, triage, and emergency decision-making~\cite {masanneck2024triage,shekhar2025use,shen2025enhancing}. AI promises enhanced pattern recognition, integration of vast data sources, real-time risk stratification, and decision support that could reduce cognitive load and improve outcomes. However, the integration of AI into dynamic, distributed work systems like EMS raises profound sociotechnical questions that extend far beyond technical accuracy.

Recent research showed that AI integration poses threats to SA across all three levels, with cascading implications for distributed team coordination. 
At level 1 SA, operators monitoring automated systems detect failures more slowly than when actively performing tasks because passive monitoring leads to attentional disengagement and reduced perceptual processing, termed  ``out-of-the-loop'' problem~\cite{casner2016challenges}. In EMS, if AI monitors vital signs or diagnostic sensors, paramedics may develop poorer Level 1 SA regarding subtle patient changes. At Level 2 SA, 
confusion arises from the opacity of the AI's ``black box.'' When EMS clinicians cannot understand what the AI is doing or why it produces specific outputs, they are unable to interpret its conclusions, calibrate appropriate trust, or detect when its assumptions conflict with ground truth~\cite{endsley2017}. Level 3 SA threats are 
critical: AI weakens EMS clinicians’ ability to project future states by replacing trajectory understanding with stepwise recommendations and by obscuring uncertainty, particularly in out-of-distribution situations~\cite{endsley2023}. 

These challenges intensify for human-AI teams, 
where effective collaboration requires managing three dimensions of awareness~\cite{endsley2023}: \textit{taskwork SA} (understanding the operational environment and system state), \textit{agent SA} (awareness of the AI system's and teammates' capabilities, limitations, and projected actions), and \textit{teamwork SA} (shared understanding of goals, roles, and dynamic task allocation).

The cognitive, social, and procedural factors currently enabling SA maintenance under extreme conditions must inform AI design if such systems are to improve rather than degrade coordination in practice. This requires moving beyond laboratory accuracy metrics and technical demonstrations to examine how experienced EMS clinicians understand SA requirements, their current technology-enabled SA practices, and their concerns about AI's impact on the coordination mechanisms through which distributed teams currently deliver effective care. Our interview study with 25 EMS clinicians across diverse organizational contexts addresses this gap, providing empirical insights into how SA functions as a coordination mechanism in current practice and how AI integration might enhance or undermine it.

\section{Study: AI Readiness and Perceptions in Emergency Medical Services}

This study examines how EMS clinicians in the United States use technology in support of their workflow in stress, high-stakes environments, and how they perceive the risks and benefits of AI-based decision support. Through semi-structured interviews with 25 EMS clinicians across municipal, private, and volunteer services, we analyze the cognitive, social, and procedural dimensions of emergency response work and attitudes toward AI. Our thematic analysis shows how EMS workflows unfolds in practice, where current technologies succeed or fail, what concerns EMS clinicians have regarding the integration of AI into their workflows and how they imagine AI tools that could support them.

\subsection{Participants}
We conducted semi-structured interviews with 25 EMS clinicians across the United States. Participants represented a range of roles, including paramedics, emergency medical technicians (EMTs), advanced EMTs, and two EMS chiefs of one of the largest fire departments in the United States. Their experience ranged from 1 to 35 years in service and encompassed urban, suburban, and rural settings. Participants’ certifications and levels of responsibility varied, reflecting the diversity of the EMS system in the U.S. Table \ref{tab:demos} provides details regarding demographics and EMS experience per role. EMS chiefs are categorized as paramedics, as both EMS chiefs hold paramedic certifications. In this paper, we collectively refer to emergency medical technicians (EMTs), advanced EMTs, and paramedics as EMS clinicians, consistent with terminology used by the US federal government 
\cite{NHTSA2025Becoming}.

\begin{table*}[h!]
\small
\centering
\caption{Demographics and EMS Experience by EMS Role. Paramedics include flight and critical care paramedics.
}
\begin{tabular}{lcccc}
\hline
\textbf{EMS Certification} & \textbf{N} & \textbf{Age Range} & \textbf{Mean Age $\pm$ SD} & \textbf{Mean EMS Experience (Years) $\pm$ SD} \\
\hline
EMT-Basic & 12 & 18--33 & 22.92 $\pm$ 4.62 & 2.42 $\pm$ 2.24 \\
EMT-Advanced & 4 & 19--31 & 23.75 $\pm$ 5.25 & 5.00 $\pm$ 2.82 \\
Paramedic & 9 & 20--71 & 39.33 $\pm$ 20.53 & 14.11 $\pm$ 14.32 \\
\hline
\end{tabular}
\label{tab:demos}
\end{table*}

To ensure a broad perspective, we recruited from municipal, private, and volunteer EMS agencies through professional networks, regional associations, and snowball sampling. All participants provided their informed consent and were compensated with a small honorarium. The study protocol was approved by our institutional review board (IRB). 

We use pseudonyms throughout to protect confidentiality, and we omit identifying details about agencies or regions. 
\subsection{Procedure}
Interviews were conducted during June and July 2025 via Zoom or in person, lasting 22–63 minutes each. 
After signing an online consent form, participants were invited to describe their professional background and reflect on their workflow, their use of technology in support of EMS work, and their attitude towards integrating AI into the EMS workflow. The interviewer followed a semi-structured protocol. 

Interviews were recorded on Zoom and transcribed by Otter.ai, and supplemented by detailed researcher field notes written immediately after each session. Reflective memos documented emerging patterns, contextual observations, and preliminary analytic insights. 
Data collection continued until thematic saturation was reached. To assess saturation, we tracked the emergence of new themes across interviews chronologically. By the 20th interview, no new themes were raised; the final five interviews reinforced and enriched existing themes without introducing new ones. At this point, we considered saturation to be reached. Our final sample of 25 EMS clinicians is consistent with established guidelines for interview-based qualitative research \cite{guest2006howmany, hennink2017saturation}. 

\subsection{Interview Guide}
The interview guide was organized around five key domains aligned with our research questions:
\begin{itemize}
 
    \item \textbf{Professional Background and Training:} Participants’ experience in EMS, education, and motivations for joining the field.
    \item \textbf{Emergency Response Workflow:} Description of the typical process from dispatch to hospital handoff, including coordination with police, fire, and hospital personnel; and challenges in information flow.
    \item \textbf{Operating Under Pressure:} Cognitive, emotional, and procedural responses to high-pressure calls; definitions of high- and low-stress situations; strategies for handling uncertainty and limited information.
    \item \textbf{Information Seeking and Technology Use:} Current tools and information sources used during calls; perceived gaps or friction points.
    \item \textbf{Perceptions of AI integration:} Awareness and prior exposure to AI tools; perceived benefits and risks; and participants’ visions for future AI systems in EMS.
\end{itemize}

The guide was piloted and refined for clarity and flow. Prompts invited participants to recount concrete incidents and reflect on their reasoning processes, such as:
\begin{quote}
    ``Can you describe a time when you had to make a quick decision under pressure?''  
    ``What kinds of information do you rely on most when things are uncertain?''  
    ``If you could design an AI assistant for your work, what would it do—and what should it never do?''
\end{quote}
\subsection{Data Analysis}
We employed a reflexive thematic analysis approach~\cite{braun2012thematic, Braun08082019} to identify patterns across the interview data and to interpret how EMS clinicians understand decision-making and AI readiness within their work contexts. Reflexive thematic analysis was chosen for its flexibility in capturing both shared meanings and nuanced, situated perspectives within complex organizational settings. 

The first author has field experience in emergency medical services. This insider perspective informed the research design, interview approach, and analytical process in several ways. Their familiarity with EMS protocols, terminology, and workflow enabled deeper probing during interviews and recognition of tacit knowledge. During analysis, this domain expertise facilitated interpretation of technical references and situated practices while requiring deliberate attention to avoid assumptions about shared understanding. To maintain analytical rigor, the research team engaged in regular discussions where the first author's experiential knowledge was treated as data to be examined alongside participant accounts rather than a single authoritative interpretation. This positioning, as both domain expert and researcher, enabled us to balance close understanding of EMS work with identifying patterns that might be invisible to EMS clinicians immersed in daily practice.

Data analysis proceeded in six iterative and overlapping stages:

\begin{enumerate}
    \item \textbf{Familiarization:} All interviews were recorded, transcribed, and reviewed for accuracy. Researchers read each transcript multiple times, noting initial impressions and recurrent concepts related to decision-making, teamwork, technology use, and AI attitude and perceptions. 
    
    \item \textbf{Initial Coding:} Two researchers independently conducted open coding. Codes were data-driven and descriptive, capturing phrases mentioned by participants. The researchers then met to discuss emergent codes, clarify definitions, and iteratively refine the codebook.
    
    \item \textbf{Developing a Codebook:} The shared codebook evolved through a combination of inductive and deductive processes. Deductive  concepts (e.g., “information seeking,” “decision making,” “AI trust”) were derived from the interview guide and previous literature, while inductive codes emerged from participant responses.
    
    \item \textbf{Iterative Coding and Consensus:} The full dataset was then recoded using the refined codebook. To ensure consistency and reliability, two coders jointly reviewed a random sample of transcripts (20\%) and compared. Discrepancies were discussed until consensus was reached, and the codebook was further refined to improve clarity and scope. Subsequently, the two coders completed coding all transcripts independently.  Intercoder reliability, based on the entire dataset, was assessed using Cohen’s $\kappa$, which yielded a coefficient of .89, indicating strong agreement. 
    
    \item \textbf{Theme Construction:} Codes were grouped into preliminary themes that represented broader patterns of meaning (e.g., Legal Concerns, Additional Clinical Support, and Education). We used affinity diagramming to physically arrange coded excerpts, identify natural groupings, and iteratively refine thematic boundaries by moving items between clusters and consolidating related themes.
    
    \item \textbf{Review and Interpretation:} The final themes were reviewed to ensure coherence, distinctiveness, and empirical grounding. Representative quotations were selected to illustrate each theme. 
\end{enumerate}

\section{Results}

\subsection{Emergency Response Workflow}
\label{sec:workflow}
Our participants confirmed the emergency response process described in prior work \cite{jensen2011paramedic}. In this work, we will use a classification based on the work of \citet{jensen2011paramedic}, breaking up the response into ten stages: the dispatch call,
the drive to the scene, arrival at the scene, the initial contact with the patient(s), assessment of the patient(s), treatment of the patient(s), departure from the scene, the drive to the hospital, the
activities at the hospital, and finally, activities after the emergency medical work has concluded.

From the perspective of an EMT, the emergency response starts with a \textit{call from dispatch}, with the location of the emergency, and some information about the nature of the emergency. The call is followed by a \textit{drive to the scene}. During the drive, EMS clinicians might review and discuss treatment options, protocols, and dosages. One paramedic explains anticipating needs for the response:``when we're going to a call, I'll look up the protocols for something that I think I'll have to use'' (P7).

Once EMS clinicians \textit{arrive at the scene} they perform a ``scene size-up,'' which ``is an overview of the scene to identify any obvious or potential hazards'' \cite{Bizjak2011EMR, Mistovich2019Prehospital}. If the scene is unsafe, EMS clinicians might wait for law enforcement to provide security and/or wear personal protective equipment, like masks, gloves, or gowns. If the scene is safe, EMS clinicians make \textit{initial contact with the patient(s)}. The goal is to determine, assess, and treat immediate life threats \cite{Mistovich2019Prehospital}.

The next stage of the response is patient \textit{assessment}, followed by patient \textit{treatment}. Here, following detailed protocols, EMS clinicians arrive at medical diagnoses and, if necessary, they start treatment with the goal of stabilizing the patient before transport to a hospital. Treatment is followed by \textit{departure from the scene} usually via ambulance. Here, treatment continues, and the patient is physically moved to the ambulance. Once in the ambulance, the \textit{drive to the hospital} commences. In this stage, the (re-)assessment and treatment continue. Once \textit{at the hospital}, EMS clinicians work with hospital personnel to transfer the patient to their care. This includes transferring relevant medical information from the emergency response. 

Finally, EMS clinicians complete the response with \textit{activities after the emergency response}. These include generating reports, learning, and activities to support EMT physical and mental health.

\subsection{Challenges Across the EMS Workflow}
\label{sec:challenges}
\subsubsection{Information Degradation Across the Emergency Response Chain} \label{subsec:degradation} 
Information quality is systematically degraded as it flows through the emergency response system.

EMS clinicians described the information flow from 911 callers through dispatch to field responders as akin to the children's game of telephone, where messages become distorted through repeated transmission. An EMT describes radio communication as ``a game of telephone. We get it from the caller and then the dispatching into us, so it could be something completely different by the time we get there than what we were told it was'' (B10). This is a systematic pattern that EMS clinicians have learned to expect (19/25).

The breakdown often begins at the very first contact with the 911 system: ``The communication breakdown happens at the very start of the call. More often than not, it's not dispatch's fault, but they usually get fed very, very incorrect information about these patients, the patients themselves'' (P2). Panicked callers, confused bystanders, or patients in distress provide incomplete or inaccurate descriptions that form the foundation for subsequent communication.

Even when callers provide clear information, the categorization systems dispatchers use can create mismatches between reality and reported conditions. As one participant described, ``The FDNY dispatchers use their own system, so you don't really know what you're walking into. Any injury, whether it's a toe injury or an amputation, comes over as 'injury major.' And then you got to read the call notes. And then the call notes aren't always exact. So there have been a lot of times where I've walked into abdominal pains and they've been full cardiac arrest'' (B3). 
The problem is compounded when information arrives through multiple channels simultaneously, creating contradictions rather than clarification: ``We'll hear something over the radio, and then the notes will say something entirely different. We'll get called to patient unresponsive, and then we get the note seconds later, and it says seizures'' (B13). Responders must then reconcile conflicting information while en route, attempting to prepare for multiple different scenarios simultaneously—itself a cognitively demanding task. Poor audio quality, rapid speech, and the cognitive demands of driving while processing verbal information (see e.g. \cite{palinko2010estimating}), mean that critical details are frequently missed or misheard: ``Unless we are paying attention to every word she's saying, which the quality isn't very good. They mumble a lot. They talk fast'' (B7). Communication infrastructure quality varies dramatically based on geography, with rural areas facing compounding challenges.

Some agencies use AI dispatch tools in an attempt to improve information flow, but these tools have limited success. One EMT states ``some programs use AI now to help fill gaps or try and interpret what people are saying on the phone... The problem with that is it ends up asking a few simple, really basic questions'' (B4). AI systems that fail to capture the complexity and nuance of emergency situations provide minimal value and may even create false confidence in information quality.

EMS clinicians face another critical information transfer point at the hospital. This handoff is particularly vulnerable to information loss because it occurs after responders have invested substantial cognitive effort in patient care and must now reconstruct and communicate everything that has happened. As one participant explained, critical information must be stored in working memory during transport and recalled during the verbal handoff: ``what's going on with the patient, what have we done so far, how did we find this patient, what's wrong with the patient, why did they call—a lot of that has to be stored in your working memory, and there's a lot of other input as well. So that's where information can be lost, or sometimes on the radio, you communicate something, but you're driving under a bridge, and the connection is bad, and it doesn't reach'' (P1). Documentation gaps between EMS and hospital providers compound the handoff challenge: ``There can be miscommunication between new EMTs and dispatchers. Certain documentation can get skipped over, and that causes a massive breakdown between the EMTs and the hospital providers'' (B8). The social dynamics of hospital handoffs can also affect information transfer quality and patient care: ``Sometimes, depending on your rapport with the nurses and doctors at the receiving facility, sometimes things will get lost. Sometimes they don't take you seriously. Sometimes you walk in, and they're like, 'waiting room' and sometimes you have to advocate for your patient just because they simply weren't listening, or you may have structured your report in a way that didn't convey the information that was needed to convey the seriousness of the complaint on scene'' (P7). 

This systematic information degradation has several con\-se\-quen\-ces: it forces EMS clinicians to develop verification strategies as they cannot trust initial information and must independently assess every situation, it creates cognitive load as responders must reconcile contradictory information streams while managing patient care, and finally, it undermines continuity of care. 

\subsubsection{Cognitive Overload and Mental Scaffolding}
Emergency scenes are inherently overstimulating environments in which responders must rapidly process multiple information streams while making consequential decisions. The cognitive demands are compounded by uncertainty. As one participant described: ``The situations are dynamic. They fall very quickly and are often complex, both in their details, in the resource allocation that you need to sort of treat your patient and get them to definitive care'' (P9). 

A critical insight that emerged challenges popular assumptions about human performance under stress (9/25). As one paramedic explained: ``There's a saying about heroes, rise to the occasion... Instead, I think we follow the level of our training. So when things are really stressful and everything's going wrong, it just comes back to let's work through the basics again, and let's make sure that I'm safe and that my partner's safe, then it's your training, it tends to be pretty straightforward'' (P9). This ``falling to the level of training'' rather than rising above it represents a fundamental principle of EMS clinician performance. Another participant articulated this explicitly: ``I believe that we all fall to the level of our training...we defer to these protocols in stressful situations'' (B10). 

Reliance on well-trained, automatic responses are essential rather than optional: ``If you go off of instinct, you can start to make mistakes, but you have your scene size up, so you have your primary impression... we just go off of protocol'' (B10). Protocols provide the cognitive framework that enables effective decision-making when cognitive resources are depleted. 

A consequence of cognitive overload and stress is tunnel vision, which can cause EMS clinicians to miss important information. As one participant described: ``I think sometimes the things that can get us in those high pressure situations... you get tunnel vision. You might miss other things where more lax environment... you could take 15 minutes to chat with Nana and get her medical history'' (A1). Tunnel vision represents a trade-off: the intense focus on critical interventions comes at the cost of broader situational awareness, which, in turn, might cause EMS clinicians to miss important secondary information that could affect care.

\subsubsection{Team Coordination Under Pressure}
EMS clinicians shared that partners must rapidly establish and execute role divisions to enable parallel processing of urgent tasks. As one participant described: ``While we're doing the call, typically, one person will do the full assessment, do the treatments. And then the next person is going to be talking to the family, getting demographic information ...'' (A2). The coordination often happens with only minimal verbal communication, relying on shared training and rapid situational assessments.

Some agencies establish consistent partnerships, in others different teams will be established on every shift. Regular partners develop sophisticated coordination: ``You develop kind of non verbal signals. So you just get to learn how your partner acts... and the way that they approach different things'' (A3). New teams actively establish communication norms: ``Talk with your partner early on in the shift. If you've never worked with them, be like, Hey, this is how I like to do things. This is how things are done'' (P2). 

For mixed-certification teams (EMT working with paramedic), coordination requires anticipating needs beyond one's own scope of practice: ``I usually work alongside paramedics, and so you kind of have to think beyond your scope, because that's where they're thinking, and you don't want to accidentally be trailing behind while they're 10 steps ahead'' (B13). This cognitive synchronization, understanding one's partner's thinking and likely next steps, requires substantial mental effort during already demanding situations. This synchronization depends on shared training that creates common mental models: ``Pretty much every level has all been taught the exact same stuff. So that generally is, you know, what's conveyed between the team members we already know'' (P10). 

When situations exceed training or individual knowledge, partners serve as immediate resources: ``If I do need more insight on what to do when it comes to making a decision, I'll ask my partner, if it's something that they can't help me out in the moment, maybe it's their first time experiencing this too, and maybe they don't have any insight to it. This is where we can also reach out to our supervisor'' (B2). This escalation pattern (partner first, then supervisor) reflects the nested coordination structure of EMS work.

Partners must balance independent judgment with trust in each other's assessments, particularly when one partner is not physically present to verify conditions. One participant described a discrepancy with their partner, stating ``I had a conflict with a paramedic not too long ago over launching a helicopter for critical patient... that was communication breakdown in part, and also lack of trust because she wasn't on scene yet. I was in a remote area and I did not have direct communication with her, so I was relying on somebody else to relay my findings to the medics that were in route, and I wanted to launch a helicopter, but the paramedic who was not on scene said not to launch'' (P9). This reveals how coordination breaks down when partners cannot directly verify each other's assessments and must rely on mediated communication.

When partners do disagree, the dynamic nature of emergencies requires rapid resolution: ``If your partner's attitude changes, you don't have to know why. You just have to get on board with the program and figure it out later'' (P9). There is no time for extended discussion; partners must quickly align even if they don't fully understand each other's reasoning.

These findings indicate that team coordination under pressure 
is based on hierarchical leadership, rapid trust calibration, tacit knowledge through repeated interaction, and shared mental models from common training. When team coordination is effective, it enables distributed cognition, and cognitive work is shared effectively across partners while maintaining awareness and coordination.

\subsubsection{Multi-Agency Coordination Complexity}
Coordination extends beyond the immediate EMS partner to include fire, police, and additional medical units. While generally effective, this multi-agency coordination is not universal: ``Obviously, like fire and PD don't know all of our policies, just like I don't know all of their policies for things, especially when it comes to the police and the actions they make'' (B1). Different organizational cultures, priorities, and procedures create friction points.
Fire departments typically assist with equipment and initial assessment: ``Sometimes it's we walk into the scene and we're not quite sure what's going on until we start talking to the fire department... Once we accept the call from them, they typically give us a report, give us a rundown'' (B9). This handoff from fire to EMS requires effective information transfer and clear transition of patient responsibility.
During high-acuity calls, the coordination demands that ``everybody knows their roles and responsibilities'' (B2). Maintaining this clarity as fire, police, additional ambulances, and potentially air medical resources converge requires substantial cognitive and communication bandwidth. 
\subsubsection {Siloed Post-Call Learning and Knowledge}
\label{subsec:siloed} Post-call debriefs vary dramatically in their existence and effectiveness. Some EMS clinicians actively seek learning opportunities: ``I always tell my partners every call, especially if it's acute, we are going to talk about this call afterwards...and do not be afraid to tell me where I could have improved, and I won't be afraid to tell you where you can improve, because that's what makes you better providers'' (P2). This commitment to mutual feedback creates structured learning moments where partners can reflect: ``You do kind of have to do the call and learn from experience and also discuss after what you can do better, what you missed, any advice that you can get, any good criticism that you can get'' (B2). However, debrief practices are far from universal: ``Some crews like to debrief. Some people don't'' (B6). Even when debriefs occur, they may be brief and unstructured, with learning captured only in individual memory rather than organizational knowledge. 

Moreover, lived experience proves essential: ``Lived experiences is just as important as maintaining the skills... training is great for the actual physical and tactile portion of medicine, or even the memorization of drug dosages and stuff, but the actual management of stress, I think, has to come from lived experience'' (P2). Yet critical lessons from unusual calls, successful improvisations, near-misses, or coping mechanisms, often remain siloed with the individuals who experienced them, or the crew they shared with, rather than becoming organizational knowledge that could benefit others facing similar situations.

\begin{figure}[h] 
    \includegraphics[width=1.0\linewidth]{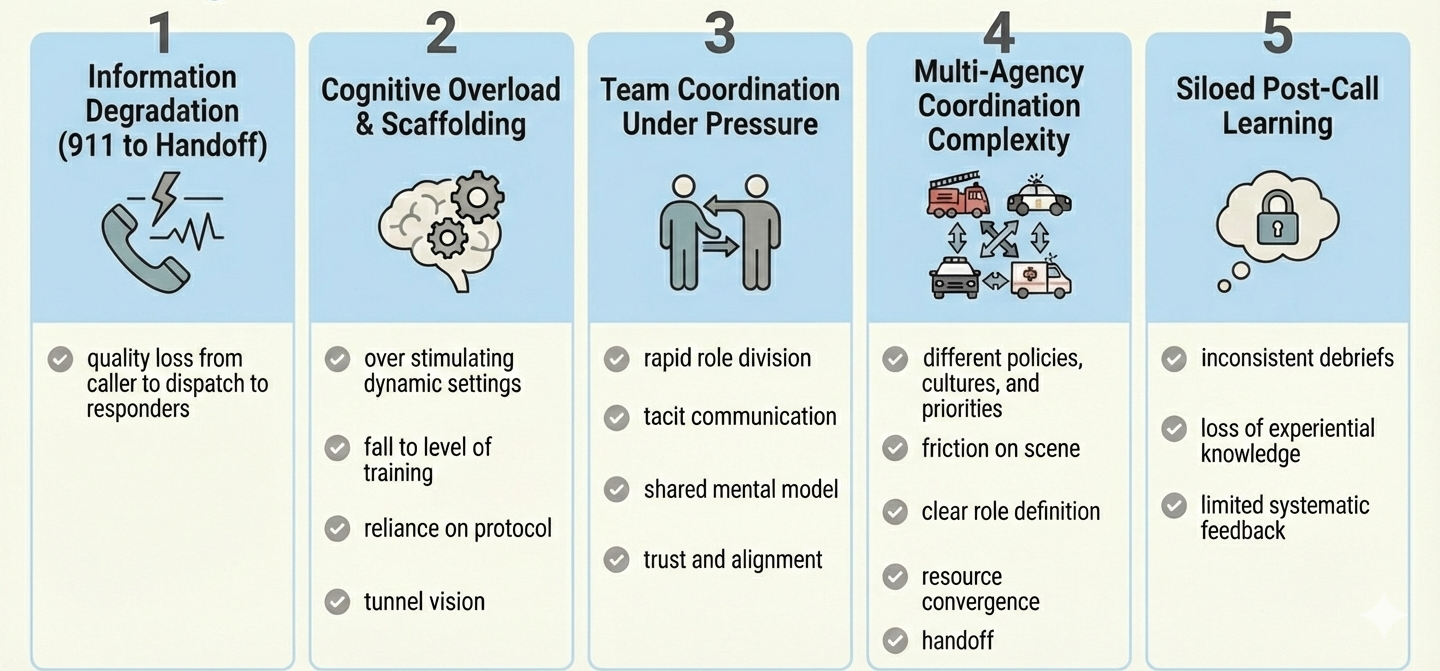}
    
    \caption{Challenges across the EMS workflow described in sections \ref{subsec:degradation} to \ref{subsec:siloed}.}
    \Description{A horizontal row of five numbered boxes, each representing one challenge in EMS workflows. Box 1, labeled ``Information Degradation (911 to Handoff),'' shows an icon of a telephone with signal waves. It lists one bullet point: quality loss from caller to dispatch to responders. Box 2, labeled ``Cognitive Overload \& Scaffolding,'' shows an icon of a brain with two gears in the foreground. It lists four bullet points: over stimulating dynamic settings; fall to level of training; reliance on protocol; and tunnel vision. Box 3, labeled ``Team Coordination Under Pressure,'' shows an icon of two people standing together with arrows pointing from one person to the other. It lists four bullet points: rapid role division; tacit communication; shared mental model; trust and alignment. Box 4, labeled ``Multi-Agency Coordination Complexity,'' shows an icon of multiple emergency vehicles including an ambulance and fire truck, and arrows crisscrossing between them. It lists five bullet points: different policies, cultures, and priorities; friction on scene; clear role definition; resource convergence; and handoff. Box 5, labeled ``Siloed Post-Call Learning,'' shows an icon of a thought bubble that encloses a padlock. It lists three bullet points: inconsistent debriefs; loss of experiential knowledge; limited systematic feedback.}
    
    \label{fig:challenges}
\end{figure}

Taken together, the five challenges described in sections \ref{subsec:degradation} to \ref{subsec:siloed} 
are interconnected aspects of a work system where information flows unreliably across organizational boundaries, cognitive resources are extremely strained, effective performance depends on distributed cognition and situational awareness, and experiential knowledge often fails to transfer beyond individual EMS clinicians. Figure \ref{fig:challenges} summarizes these challenges.

\subsection{Current Use of Technology}

Our semi-structured interviews with participants included questions about technology use. We created a list of technologies that they mentioned, leaving out manufacturer and product names, and identifying devices by their functionality. E.g., ``Toughbook'' is the product name of a popular rugged laptop, however, this name is often used to denote all rugged laptops. We converted all mentions of ``Toughbook'' to ``tablet or laptop.'' Next, for each technology, we assessed how EMS clinicians can interact with it, and which emergency response stages they are used in. The result is Table \ref{tab:technologies}.

The first column of Table \ref{tab:technologies} lists the technologies that our participants mentioned. These range from general-purpose devices that have been common for decades (such as the radio), to specialized medical tools (such as the automatic CPR device), to AI-based software. 

The next four columns use Wickens' multiple resource theory \cite{wickens2008multiple} to describe how EMS clinicians interact with the technology. To perceive the information provided by the devices, EMS clinicians use visual (Vis.) and auditory (Aud.) resources. They control the technology with manual actions (Man.) or with speech (Spe.). We identified the input modalities and output activity that EMS clinicians need to interact with each technology using our understanding of how each technology
works. For example, the radio often has a visual display to indicate which channel it is tuned to, and it produces audio that the user listens to. The radio also has a push-to-talk button, which requires manual resources to operate, and the EMS clinician speaks into the radio, which requires speech resources. It is interesting to note that for almost all of the technologies we list, the EMS clinician will consume visual information, and they will use manual control to operate the technology.

The next ten columns represent the ten response stages of a team of EMS clinicians \cite{jensen2011paramedic}: the dispatch call (Disp.), the drive to the scene (Drive), arrival at the scene (Arrive), the initial contact with the patient(s) (Con.), assessment (Assess) and treatment (Treat) of the patient(s), departure from the scene (Depart), the drive to the hospital (To H), the activities at the hospital (At H), and finally activities after the emergency medical work has concluded (After). We see that the radio, and the mobile phone, are used throughout the response, an automatic CPR device or a ventilator is used once treatment starts, and reporting tools are used towards the end of the response. Our interviews also reveal that although AI tools are rarely used, they are sometimes used as early as the treatment stage, and that they are certainly used after the emergency part of the response, when EMS clinicians review the events, and work to improve their clinical knowledge.


\begin{table*}[h!]
  \centering
  \caption{Technologies used by EMS Clinicians. The table lists the technologies and indicates
    which resources are needed to interact with each technology (visual and/or auditory
    perception, manual and/or speech response). The emergency response stages where the
    technology is commonly used is indicated.}
  \label{tab:technologies}
  \setlength{\tabcolsep}{4pt}
  \begin{tabular}{P{3.8cm}|cccc|cccccccccc}
    \toprule
    & \multicolumn{4}{c|}{\textbf{Resources}}
    & \multicolumn{10}{c}{\textbf{Response stage}} \\
    \textbf{Technology}
      & Vis. & Aud. & Man. & Spe.
      & Disp. & Drive & Arrive & Con. & Assess & Treat & Depart & To~H & At~H & After \\
    \midrule
    Radio                          & x & x & x & x & x & x & x & x & x & x & x & x & x & x \\ \hline
    Mobile data terminal (MDT)     & x & x & x & x & x & x & x &   &   &   &   & x &   &   \\ \hline
    Mobile phone voice             &   & x &   & x & x & x & x & x & x & x & x & x & x & x \\ \hline
    Mobile phone text              & x &   & x &   & x & x & x & x & x & x & x & x & x & x \\ \hline
    Mobile phone dispatch app      & x &   & x &   & x &   &   &   &   &   &   &   &   &   \\ \hline
    Mobile phone navigation        & x &   & x &   &   & x &   &   &   &   &   & x &   &   \\ \hline
    Online search                  & x &   & x &   &   &   & x &   & x & x & x & x & x & x \\ \hline
    Language translator apps       & x & x & x & x &   &   &   & x & x & x & x & x & x &   \\ \hline
    Monitor and defibrillator      & x & x & x &   &   &   &   &   &   & x & x & x & x &   \\ \hline
    Automatic CPR device           & x & x & x &   &   &   &   &   &   & x & x & x & x &   \\ \hline
    Ventilator                     & x &   & x &   &   &   &   &   &   & x & x & x & x &   \\ \hline
    Medical support apps           & x &   & x &   &   & x &   &   & x & x & x & x &   &   \\ \hline
    MD consultation phone line     &   & x &   & x &   &   &   &   & x & x & x & x &   &   \\ \hline
    Auto lifting/loading stretcher & x &   & x &   &   &   & x &   &   &   & x &   &   &   \\ \hline
    Tablet or laptop               & x &   & x &   &   &   &   &   & x & x &   & x & x & x \\ \hline
    Reporting tools                & x &   & x &   &   &   &   &   & x & x &   & x & x & x \\ \hline
    General-purpose AI tools       & x &   & x &   &   &   &   &   &   &   &   &   &   & x \\ \hline
    Specialized AI tools           & x &   & x &   &   &   &   &   &   & x & x & x &   & x \\
    \bottomrule
  \end{tabular}
\end{table*}

\subsection{The Technologies That EMS Clinicians Use}
\label{tech-emt-use-details}
Throughout the response stages, EMS clinicians use multiple technologies for communication with remote members of the team. These include the \textit{radio}, the \textit{mobile data terminal (MDT)} (which is an in-vehicle computer used by first responders \cite{kun2005computers}), the voice and text features of the \textit{mobile (or cell) phone}, and the \textit{mobile phone dispatch app} which is used to send dispatch messages to EMS clinicians. 

EMS clinicians also use commercial apps including for \textit{navigation}, \textit{online search}, and \textit{language translation}. The latter can be used to communicate with patients.

While performing medical tasks, EMS clinicians use an array of medical devices. These include devices that combine medical \textit{monitors} (e.g. for blood pressure and oxygenation), which includes a \textit{defibrillator}. A relatively new device is the \textit{automatic CPR device}, which makes it possible to continue with CPR even while the patient is physically moved. \textit{Ventilators} help patients breathe. 

When EMS clinicians need advice, they can turn to a variety of \textit{medical support apps}, or \textit{consult by phone with a medical doctor (MD)}. When EMS clinicians are ready to load the patient into the ambulance, they can take advantage of 
the \textit{auto lifting/loading stretcher}, 
which reduces lifting-related injuries. 

During the response, EMS clinicians often use hardened \textit{tablets or laptops}, which can withstand physical impacts, and are bright enough to be used outdoors in daylight. They might access their \textit{reporting tools} (software) through these devices.

Finally, EMS clinicians use AI tools, albeit infrequently. They use \textit{general-purpose AI tools} after the patient is in the hospital, to improve their understanding of disease processes, to review  diagnoses from the call, to develop training scenarios, and to generate reports. They also use \textit{specialized AI tools} to support their clinical work and as follow-up after the clinical work is completed. The specialized tools are used to rapidly review medical literature. 

\subsection{Current Use, Concerns, and Envisioned Applications of AI}
\label{sec:ai-use-and-concerns}
To understand EMS clinicians' perceptions and attitudes towards AI integration in EMS workflows, we first asked participants about their current use of AI and its perceived benefits. Participants shared from their experience using AI for both work-related and non-work related tasks. Most of the EMS clinicians (19/25) mentioned the use of AI in an educational context. Such usage includes AI tutoring, generating scenarios, assistance in writing documents and emails, programming, and summarizing information. For example, one advanced EMT participant described that ``in the training environment ..., we've used AI to help create patient scenarios to use as training opportunities for Cadet members or new EMTs'' (B12). Some participants (6/25) use AI for EMS work-related tasks including writing documentation and reports, and learning about a topic. For example, one paramedic mentioned that ``Open Evidence, AI is really good about having some sort of peer reviewed journal or article that's attached to it, or a link, like a hyperlink that's in reference to whatever information it's spitting out'' (P10). Several EMS clinicians (7/25) mentioned direct benefits related to saving time and reducing cognitive load. For example, one participant shared that ``you could ask AI many, many questions and get a lot of hours worth of information sunk into a couple paragraphs'' (A1).

Participants expressed diverse concerns about the introduction of AI into EMS workflows, including how AI might alter EMS clinicians' autonomy, legal issues, patient outcomes, and the lived realities of work in high-stakes, time-critical environments.

\subsubsection{Legal and Privacy} \label{subsec:laws}

Legal and privacy-related concerns were also prominent. EMS clinicians expressed uncertainty about how AI systems would handle sensitive patient data and comply with regulations such as HIPAA (13/25). Some participants were also concerned about data security and protecting data from unauthorized access (7/25) - ``with AI, everything is training data, everything is whatever. It is never secure'' (B7).

The questions of assigning liability, especially in cases where AI-supported actions result in adverse outcomes, 
were also mentioned by participants (9/25). In the words of P11, ``AI says, Hey, use this protocol. This is what's going on. But if it's wrong, I'm still the one who's, you know, held liable for it.''

\subsubsection{Technical integrity: Accuracy, Bias, and Reliability}
EMS clinicians expressed major concerns about AI accuracy and hallucinations (19/25) as well as nuanced concerns about how AI systems might interact with human error and data quality in emergency contexts. Participants emphasized that AI could amplify, obscure, or redistribute errors if it relies on incomplete, inaccurate, or context-insensitive data (7/25). As one EMT describes, ``sometimes it doesn't understand context, especially if there was audio or video feed, bad lighting, bad audio, combative scene'' (B10). Participants also noted that inaccurate data or erroneous data entry could lead to downstream error (10/25). For example, one paramedic explained that ``it may be that you entered the information in a misleading way, or you didn't.... you left out a vital, or you put a vital in there, or you emphasize something, or you ask the question a certain way. I don't know the AI is easily misled, but I think it can only spit back once you've given it really solid facts'' (P5).

EMS clinicians also worried about bias (9/25), fearing that biased outputs could override situational judgment, especially for less experienced EMS clinicians. For example, P1 noted that ``medicine is a very biased field... and that data that we get to train those AI are also skewed and biased... in terms of bias, here's one big example, heart attacks [in] women. Heart attack signs and symptoms were studied on white males, or males, in particular, men. The [symptoms of] clutching chest pain, the chest pressure, the radiating to the left side, ... that was studied in males. That's so, so, so unfair for women... there's a study out there, on PubMed, that states that... cardiac pathologies [in women] were misdiagnosed as anxiety by doctors.'' Finally, concerns about technological reliability were common (7/25), particularly regarding connectivity and system failure in rural or low-infrastructure settings. As one EMS clinicians put it, ``If it loses connection out in the field... no one’s ever going to take it out again, because they’re not going to trust it'' (B13).

\subsubsection{Human-Centric Limits}

EMS clinicians emphasized that AI systems lack the human judgment, contextual sensitivity, empathy, and embodied situational awareness required for emergency medical work (13/25). Participants described EMS practice as relying on subtle, situational cues such as patient demeanor, scene safety, and team dynamics, that are difficult to capture in data, leading to skepticism about AI-generated recommendations in the field. Participants highlighted the importance of tacit knowledge and EMS clinician's intuition in the field that is lacking from AI systems. In the words of one participant, ``there's always going to be something missing from a physical person not being able to interact with this patient, whether it's... stepping into that environment and you kind of get certain feelings that you know an AI might not be able to understand or think about, or they might not have the context that you have'' (B1). 

\subsubsection {Threats to Professional Autonomy and Expertise}

A central concern (11/25) involved the loss of professional autonomy and control. Participants worried that AI systems could constrain clinical judgment by prescribing actions, narrowing decision pathways, or prioritizing algorithmic recommendations over experiential knowledge. In the words of one paramedic: "I again, personally recoil from it, just because that's, it's taking all of the thinking out of it, you know, then we're just bodies that move other bodies around" (P7). This concern was closely tied to fears of AI replacing, devaluing, and eroding EMS expertise, particularly if systems were adopted as cost-saving substitutes rather than as supportive tools. For example, one paramedic shared: ``it could potentially make providers complacent, in a way, just because then they're relying very heavily on the AI instead of actually doing the clinical judgment for themselves, which I know is kind of the point of that, but it also ... dangerous, especially for patient care outcome'' (P2). Many EMS clinicians (9/25) specifically stressed the risk of over-reliance and skill erosion, especially in less experienced EMS clinicians. 
A few participants (2/25) raised concerns about surveillance and micro-management, especially if AI were used to monitor performance, enforce compliance, or retrospectively evaluate decisions made under extreme time pressure.

\subsubsection{Workflow Friction} \label{subsec: workflow}
EMS clinicians expressed concern that AI systems could increase cognitive load and disrupt attention during already demanding emergency medical workflows (15/25). 
Some participants (2/25) emphasized the risk of distraction from patient care, expressing concern that interaction with AI interfaces could divert attention away from patients and interpersonal engagement. One EMT explained, ``If I’m with a patient, I like to be able to engage with the patient. I don’t want to be talking to the patient while staring at my phone... asking AI to add up all of these symptoms'' (B12). Time constraints further shaped skepticism toward AI use in the field. EMS clinicians (4/25) described AI interactions as potentially too time-consuming, especially when outputs must be verified against other trusted sources. 

\begin{figure}[h] 
    \includegraphics[width=1.0\linewidth]{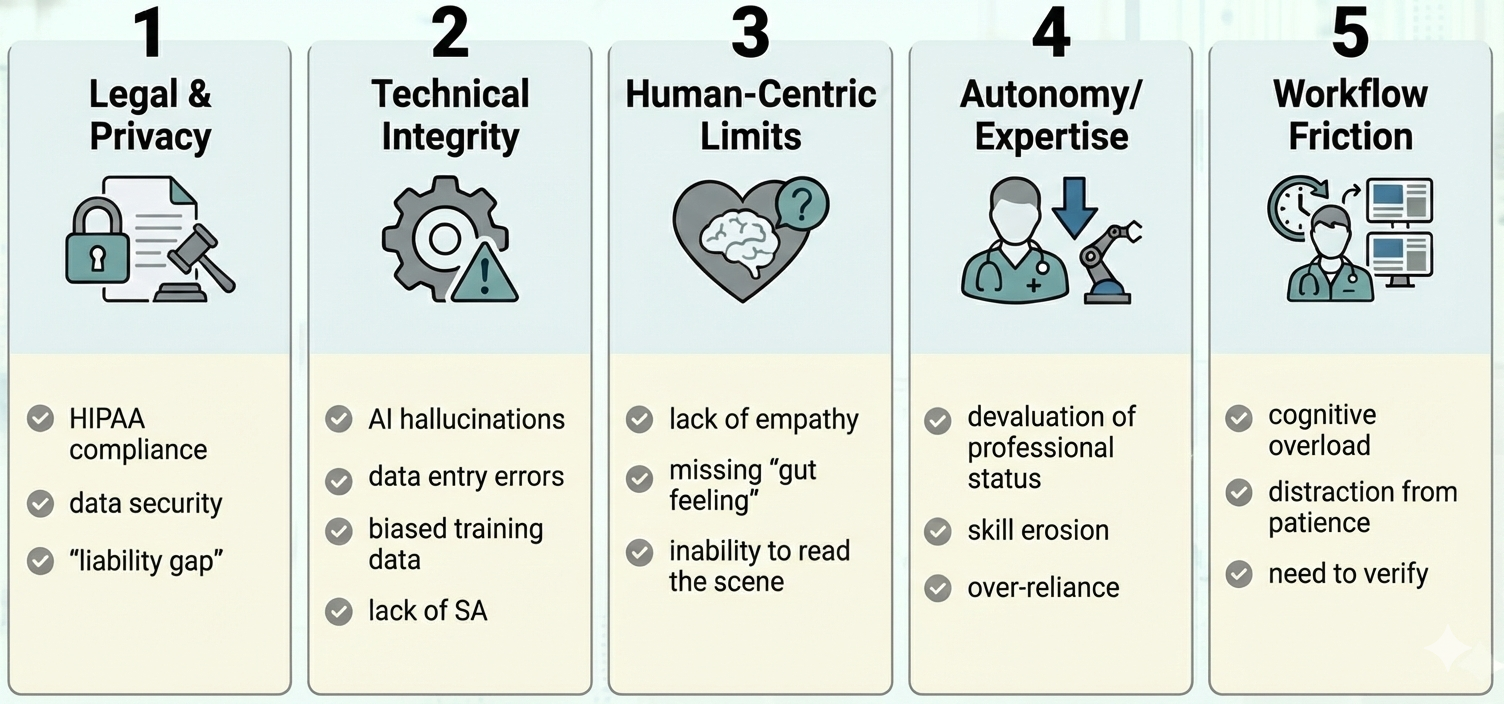}
    \Description{A horizontal row of five numbered boxes, each representing one core tension in integrating AI into EMS workflows. Box 1, labeled ``Legal \& Privacy,'' shows an icon of a lock and a document. It lists three bullet points: HIPAA compliance; data security; and ``liability gap.'' Box 2, labeled ``Technical Integrity,'' shows an icon of a gear with an exclamation point in a triangle. It lists four bullet points: AI hallucinations; data entry errors; biased training data; and lack of SA (situational awareness). Box 3, labeled ``Human-Centric Limits,'' shows a human brain enclosed in a heart-shaped icon, and a question enclosed in a circle. It lists three bullet points: lack of empathy; missing ``gut feeling''; and inability to read the scene. Box 4, labeled ``Autonomy/Expertise,'' shows an icon of a clinician with a stethoscope next to a robotic arm, with a downward-pointing arrow between them. It lists three bullet points: devaluation of professional status; skill erosion; and over-reliance. Box 5, labeled ``Workflow Friction,'' shows an icon of a clinician with a stethoscope in the foreground, as well as a clock and two computer monitors in the background. It lists three bullet points: cognitive overload; distraction from patient; and need to verify.}
    \caption{Core tensions of integrating AI into EMS workflows, described in sections \ref{subsec:laws} to \ref{subsec: workflow}.}
    
    \label{fig:tensions}
\end{figure}

Taken together, the five concerns described in sections \ref{subsec:laws} to \ref{subsec: workflow} indicate that EMS clinicians perceive significant tensions for the integration of AI to EMS workflows. Figure \ref{fig:tensions} summarizes the emerging core tensions of integrating AI into EMS work.

\subsubsection{Potential Use of AI in EMS}
Our semi-structured interviews with participants included questions about the potential use of AI in their work. We list their suggestions in Table \ref{tab:ai-suggestions}. In this table, just like in Table \ref{tab:technologies}, we denote the resources that would likely be used to interact with the suggested AI-based technology, as well as which stage of the response the suggested idea would be used in. Our identification of resources and stages is based on a combination of results from the interviews, and our own expertise. 


\begin{table*}[h!]
  \centering
  \caption{Suggested uses of AI. The table lists suggested uses, the resources needed to interact
    with the technology (visual/auditory perception, manual/speech response), and the stages of
    the emergency response where the suggested use would occur.}
  \label{tab:ai-suggestions}
  \setlength{\tabcolsep}{4pt}
  \begin{tabular}
  {P{3.8cm}|cccc|cccccccccc}
    \toprule
    & \multicolumn{4}{c|}{\textbf{Resources}}
    & \multicolumn{10}{c}{\textbf{Response stage}} \\
    \textbf{Suggested use of AI}
      & Vis. & Aud. & Man. & Spe.
      & Disp. & Drive & Arrive & Con. & Assess & Treat & Depart & To~H & At~H & After \\
    \midrule
    Transcribe speech            & x & x & x & x & x & x & x & x & x & x & x & x & x &   \\ \hline
    Support recording events     & x & x & x &   & x & x & x & x & x & x & x & x & x &   \\ \hline
    Support following protocols  & x & x &   &   &   &   &   & x & x & x & x & x & x &   \\ \hline
    Provide medical information  & x & x & x & x &   &   &   &   & x & x &   & x &   &   \\ \hline
    Discuss clinical work        & x & x & x & x &   &   &   &   & x & x &   & x &   &   \\ \hline
    Support data interpretation  & x &   & x &   &   &   &   &   & x & x &   & x &   &   \\ \hline
    Support data collection      & x & x & x &   &   &   &   &   & x & x &   & x &   &   \\ \hline
    Dose calculations in context & x &   & x &   &   &   &   &   & x & x &   & x &   &   \\ \hline
    Support diagnosis            & x &   & x &   &   &   &   &   & x & x &   & x &   &   \\ \hline
    Check human work             & x & x &   &   & x & x &   &   & x & x & x & x & x & x \\ \hline
    Support report generation    & x & x & x & x &   &   &   &   &   &   &   &   & x & x \\
    \bottomrule
  \end{tabular}
\end{table*}

As shown in Table \ref{tab:ai-suggestions}, participant suggestions for AI-based technologies can be organized into 11 groups:

\textbf{Transcribe speech.} EMS clinicians talk to patients and each other throughout the stages of the response. Furthermore, as shown in Table \ref{tab:technologies}, the radio and the voice function of the mobile phone, are used throughout the response stages. Thus, a great deal of information is exchanged verbally, and retaining this information is important for the quality of care (e.g. to inform hospital staff about issues), and for generating reports. Yet, speech is ephemeral; a tool that transcribes speech can create a permanent record of what was said.

\textbf{Support recording events.} The timing of actions is an important aspect of patient care reports. However, EMS clinicians undertake numerous actions during a response, and importantly, these actions happen rapidly and under a great deal of stress. Jensen et al. point out that EMS clinicians make many decisions per unit time (high ``decision density''). These decisions result in actions, and the timing of these actions is often important to track. A tool that automates recording the timing of at least some actions would be helpful in creating subsequent reports. 

\textbf{Support following protocols.} Given that the decision density is high, protocols are one key tool that allows EMS clinicians to perform their jobs -- without following protocols they could easily forget to perform critical steps. But, under pressure, and due to interruptions, it might sometimes be difficult to remember protocols that are rarely used (e.g. a protocol related to pediatric medication). A tool that keeps track of completed steps and presents next steps as needed, could help.

\textbf{Provide medical information.} EMS clinicians often have medical questions that can be resolved within a single turn, such as yes/no questions. Currently, these questions might be resolved with an online search (listed in Table \ref{tab:technologies}), but a dedicated tool could help improve reliability and reduce the time needed to get an answer.

\textbf{Discuss clinical work.} EMS clinicians might also have questions that take multiple turns to resolve. Currently, these might require talking to a remote medical doctor (MD) (e.g. using the ``MD consultation phone line'' from Table \ref{tab:technologies}). Accessing a remote MD 
can be delayed by connectivity issues and communication barriers, and a dedicated tool could reduce the time needed to get an answer.

\textbf{Support data interpretation.} EMS clinicians routinely read and interpret data (e.g. EKGs) as part of patient care. When complex data are involved, additional data interpretation support can be helpful. A data interpretation tool can aid EMS clinicians during high stress and rapidly-evolving situations. 

\textbf{Support data collection.} Collecting data in the uncontrolled environment of the EMS clinicians' work can be challenging. E.g., a photograph might work well for a diagnostic tool only if it is taken under certain light conditions. An AI tool could help EMS clinicians optimize data collection variables (such as lighting for a photo), to support the work of diagnostic tools.

\textbf{Dose calculations in context.} Many diagnostic and treatment actions require customized calculations. While these calculation are relatively simple, they can be challenging to perform under pressure or when an EMS clinician is tired as they might require variables such as age, weight, and sensitivity. An AI tool could support performing these calculations.

\textbf{Support diagnosis.} EMS clinicians deal with a broad variety of medical situations under time pressure and outside the organized environment of a hospital. They use their experience to arrive at medical diagnoses, but this process is made difficult by the time pressure and hectic environment. An AI tool could support the diagnostic process by relying on a broader corpus of experiences, and by providing diagnosis hypotheses to the EMS clinician.

\textbf{Double-check human work.} EMS clinicians complete long and stressful shifts, and often work with few hours of sleep. Under these conditions, they could benefit from a tool that can double-check the medication type and dose they are about to administer. (Remember that we asked EMS clinicians to imagine a 100\% accurate AI.) 

\textbf{Support report generation.} Report generation is important, but also a time-consuming and draining task that EMS clinicians do not consider to be professionally rewarding. A tool that automates parts of the report generation process could reduce the time required for this task.

\section{Discussion}
We conducted semi-structured interviews with 25 EMS clinicians to examine technology use, workflow challenges, and AI perceptions in emergency medical practice. Here, we discuss how our findings connect to theories of distributed cognition and situational awareness, identify core tensions regarding AI integration, and derive design implications for human-centered AI in high-stakes, time-critical emergency response

\subsection{Distributed Cognition and SA Under Constraint}
Our study reveals that emergency medical work fundamentally depends on distributed cognition across partners, agencies, and artifacts under conditions of extreme cognitive load, unreliable information flow, and severe time pressure. We identified five interconnected challenges (Figure \ref{fig:challenges}): information degradation across the emergency response chain, cognitive overload requiring protocol scaffolding, distributed team coordination under pressure, multi-agency coordination complexity, and siloed post-call learning. These challenges demonstrate that effective emergency response emerges not from individual decision-making, but from successfully managing both distributed intelligence across people and systems \cite{hutchins1995}, and team situational awareness \cite{endsley_jones2001}.

Interestingly, we find that EMS clinicians describe their actions in highly stressful situations in a manner similar to firefighters. 
Klein~\cite{klein1993recognition} found that experienced fireground decision makers rapidly ``recognized'' situation types and knew ``the typical way of reacting to it''~\cite{klein1993recognition}, a process that bypasses option comparison altogether. Our findings show identical mechanisms in EMS, as reliance on well-trained, automatic responses was described as essential rather than optional: ``If you go off of instinct, you can start to make mistakes, but you have your scene size up, so you have your primary impression... we just go off of protocol'' (B10). This suggests that EMS clinicians, like Klein's fireground commanders, rely on recognition-primed decision processes where situation awareness and assessment trigger immediate, protocol-based action rather than deliberate option evaluation.


\subsection{Current Technology Use}
Our analysis of current technology use (Table \ref{tab:technologies}) reveals a critical resource competition problem that directly threatens SA: nearly all technologies require visual and manual resources across the stages of the response. Thus, technology interaction competes with the perceptual activities (Level 1 SA) necessary for detecting patient vital signs, environmental hazards, and partner actions. Radio and mobile phone ubiquity across all response stages reflects the success of these communications technologies in supporting situational awareness, as they do not demand continuous visual attention. However, the systematic information degradation EMS clinicians described, the ``game of telephone'' from 911 through dispatch to field (sometimes due to radio quality), undermines Level 1 SA by providing unreliable perceptual input, forcing EMS clinicians to develop verification strategies that consume cognitive resources. Similarly, the field-to-hospital handoff, which relies on EMS clinicians' working memory, represents a Level 2 SA failure when EMS clinicians cannot maintain accumulated patient information under competing demands, leading to information loss. Critically, we found that EMS clinicians ``fall to the level of training'' rather than rise to occasions, meaning protocol scaffolding serves as essential external cognitive structure enabling Level 3 SA (projection). When internal cognitive resources are depleted, EMS clinicians can anticipate next steps because protocols provide predictable sequences. 

These findings align with Wilson et al.'s \cite{wilson2023scoping} emphasis on understanding information trajectories in distributed acute care work and Endsley's \cite{endsley_jones2001, endsley2015} framework showing that SA quality fundamentally determines decision quality, while revealing the profound challenge of maintaining SA across distributed teams when cognitive resources are overwhelmed and information trajectories are unreliable.

\subsection{Core AI Tensions}
We identified five core tensions regarding AI integration (Figure \ref{fig:tensions}): legal/privacy concerns, technical integrity issues, human-centric limits, threats to autonomy and expertise, and workflow friction. These tensions affect how AI systems could undermine the situational awareness, EMS clinician judgment, and coordination mechanisms that effective emergency response requires. These concerns map directly onto Endsley's \cite{endsley2017, endsley2023} theoretical framework for AI's impact on situational awareness across all three levels: EMS clinicians fear that lack of sensitivity to bad lighting, bad audio, multiple actors on the scene, will degrade Level 1 SA (perception of critical cues during already-problematic tunnel vision), lack of contextual sensitivity opacity will prevent Level 2 SA (comprehension of situation meaning, particularly when AI lacks understanding of subtle cues, and patient demeanor), and AI recommendations will undermine Level 3 SA (projection of future states) by replacing trajectory understanding with stepwise suggestions detached from protocol frameworks. 

Notably, EMS clinicians emphasized threats to professional autonomy and skill erosion, fearing AI will create ``bodies that move other bodies around'' (P7) rather than experts. This aligns with automation-induced performance degradation research \cite{casner2016challenges}. This also connects to the ``fall to the level of training'' principle: if AI replaces protocol-driven practice, EMS clinicians might lose the skills they would need to rely on when AI fails. Research on time-pressured AI-assisted decision-making \cite{cao2023time, jacobs2021designing, swaroop2024accuracy} shows that limited observation time increases reliance. EMS clinicians worried that EMS' extreme time pressure would push uncritical dependence, especially among novices. 

Concerns about biased training data (e.g., gendered symptoms ``misdiagnosed as anxiety'') \cite{{hou2025equitable}}, infrastructure reliability, and legal issues highlight that explainability and transparency, trustworthiness, and accountability \cite{kun_accountable_ai_forthcoming} are critical factors in the deployment of AI for EMS.

\subsection{Effects of Suggested Uses of AI}
\subsubsection{A focus on emergency care subtasks}

All but one suggested use of AI presented in Table \ref{tab:ai-suggestions} would support work during three phases of the response where EMS clinicians provide emergency medical care: during assessment, treatment, and the drive to the hospital (when treatment continues). EMS clinicians are focused on the hands-on tasks that save lives, and this focus is reflected in their suggested uses of AI. 

Table \ref{tab:ai-suggestions} shows that EMS clinicians rarely suggested AI for the two initial stages on the scene of the response: the arrival at the scene, and the initial contact with the patient(s). This is not surprising, since today's AI would struggle with assessing the context during the arrival stage. Furthermore, the initial contact with the patient is a deeply consequential stage of the response where EMS clinicians must be fully in charge.

And, despite significant concerns about using AI, EMS clinicians envision stage-appropriate roles for AI that address the resource competition problem revealed in Table \ref{tab:technologies}. 
Namely, Table \ref{tab:ai-suggestions} shows that many of the suggested AI applications could utilize auditory perception, instead of, or in addition to, visual perception. This could reduce competition for visual perception, which is critical for maintaining situational awareness, and for working with patients. 





\subsubsection{Broader effects of using AI}
Inspired by the work of Acemoglu and Johnson \cite{acemoglu_johnson_power_progress_2023}, in Table \ref{tab:ai-suggestions-effects} we evaluate the possible effects of the suggested uses of AI presented by our participants. Table \ref{tab:ai-suggestions-effects} provides one key message: \textit{all} of the suggested uses of AI would improve the quality of EMS clinician work, while the effects on time reallocation, new tasks for EMS clinicians, job losses for others, and EMS clinician well-being is modest. We see that EMS clinician suggestions focus on benefits to their patients.

\begin{table*}
  \centering
  \caption{Possible effects of suggested uses of AI. Cells marked in green denote desirable
    outcomes (such as improved quality of work). Cells marked in pink denote undesirable
    outcomes (such as job loss for other workers).}
  \label{tab:ai-suggestions-effects}
  \setlength{\tabcolsep}{4pt}
  \begin{tabular}{P{4cm}|P{1.4cm}|P{1.4cm}|P{1.6cm}|P{1.6cm}|P{1.6cm}|P{1.4cm}|P{1.6cm}}
    \toprule
    \textbf{Suggested use of AI}
      & \textbf{Challenges of EMS workflow (sect.~4.2) addressed?}
      & \textbf{Core AI tensions (sect.~4.5) present?}
      & \textbf{Support reallocating time toward work with patients?}
      & \textbf{Support improved quality of work?}
      & \textbf{Support new, high-value tasks for clinician?}
      & \textbf{Possible job loss for others?}
      & \textbf{Support improved clinician well-being?} \\
    \midrule
    Transcribe speech            & 1    & 1, 2          & No                        & \cellcolor{cellgreen}Yes & No                        & No                        & No \\ \hline
    Support recording events     & 2    & 1, 2          & No                        & \cellcolor{cellgreen}Yes & No                        & No                        & No \\ \hline
    Support following protocols  & 2    & 1, 4, 5       & No                        & \cellcolor{cellgreen}Yes & No                        & No                        & No \\ \hline
    Provide medical information  & None & 1, 2, 3, 4, 5 & No                        & \cellcolor{cellgreen}Yes & No                        & No                        & No \\ \hline
    Discuss clinical work        & None & 1, 2, 3, 4, 5 & No                        & \cellcolor{cellgreen}Yes & No                        & \cellcolor{cellpink}Yes  & No \\ \hline
    Support data interpretation  & None & 1, 2, 3, 4, 5 & No                        & \cellcolor{cellgreen}Yes & \cellcolor{cellgreen}Yes  & No                        & No \\ \hline
    Support data collection      & 1    & 1, 2, 3, 4, 5 & No                        & \cellcolor{cellgreen}Yes & No                        & No                        & No \\ \hline
    Dose calculations in context & 2    & 1, 2, 4       & No                        & \cellcolor{cellgreen}Yes & No                        & No                        & No \\ \hline
    Support diagnosis            & None & 1, 2, 3, 4, 5 & No                        & \cellcolor{cellgreen}Yes & No                        & \cellcolor{cellpink}Yes   & No \\ \hline
    Check human work             & 2    & 1, 4, 5       & No                        & \cellcolor{cellgreen}Yes & No                        & No                        & No \\ \hline
    Support report generation    & None & 1, 2          & \cellcolor{cellgreen}Yes  & \cellcolor{cellgreen}Yes & No                        & No                        & \cellcolor{cellgreen}Yes \\
    \bottomrule
  \end{tabular}
\end{table*}

\subsection{Implications for Design}
Our findings reveal five critical design principles for AI systems that alleviate challenges in EMS workflows by enhancing rather than undermining the distributed cognition, situational awareness, and professional judgment that effective emergency response requires.
\subsubsection{Support Distributed Cognition, Don't Centralize It} Current EMS work distributes cognitive work across partners, organizational boundaries, and artifacts. AI systems should enhance this distributed intelligence, not collapse it through centralized recommendations.
AI interfaces should be designed for team-level SA by providing shared information spaces. For example, visualizing patient vitals trajectories can serve as reference for the team rather than alert individuals separately \cite{endsley_jones2001}. 
Role-based information should be accessible without creating silos: EMTs and paramedics need different information depths, but protocol support should show EMT-level basics with expandable paramedic-level options so both see what information exists at different expertise levels. 
Finally, designers should preserve multi-channel communication diversity (radio, phone, MDT) rather than seek to consolidate to single AI interface. Hutchins \cite{hutchins1995} showed redundancy in communication channels provides resilience essential for coordination under stress.
\subsubsection{Enhance Situational Awareness Across All Three Levels}
AI systems risk degrading SA through passive monitoring (Level 1), opacity (Level 2), and uncertainty obscuration (Level 3) \cite{endsley2017, endsley2023}. To support Level 1 SA (perception), designers should reduce rather than increase perceptual demands: use auditory output with visual redundancy (keeps hands and eyes on patient), implement ambient sensing to reduce manual entry, and set proactive alerts for overlooked elements when tunnel vision is likely. To support Level 2 SA (comprehension), AI should show the clinical pathway behind its assessment (e.g. ``Elevated lactate + hypotension + fever consistent with sepsis''), with uncertainty visualization showing AI confidence and data quality indicators. 
To support Level 3 SA (projection) interfaces should show vital signs trends to enable trajectory understanding rather than snapshots, and reinforce protocol-based next steps rather than novel recommendations. 

\subsubsection{Preserve and Enhance Professional Autonomy}
Addressing concerns about judgment constraints, skill erosion, and EMS clinicians becoming ``bodies that move other bodies around'' (P7), designers should position AI as consultant not authority using recommendation framing with clinical rationale allowing override based on situational knowledge. AI tools should support skill development through training mode with AI guidance during low-acuity calls, performance mode with minimal AI during emergencies, AI withdrawal as experience grows. It is critically important to maintain transparency about AI capabilities, limitations, training data biases, and complete audit trails. 
Designers should consider how to support peer and supervisory oversight for learning not surveillance, and to aggregate performance insights for organization-level improvement. EMS clinicians should be able to control data sharing to protect against micro-management concerns.

\subsubsection{Adapt to Cognitive Resources Across Workflow Stages} 
Cognitive resource availability varies dramatically across stages; Table \ref{tab:ai-suggestions} reveals EMS clinicians rarely envision AI for on-scene Assessment/Treatment when cognitive load peaks. One possibility for designers is to implement acuity-responsive behavior: high-acuity (cardiac arrest, trauma) transitions AI to purely passive data collection with zero prompts requiring evaluation, medium-acuity provides optional support accessible when capacity permits, low-acuity enables active AI suggestions. Acuity could be detected automatically from call type, vitals, and scene factors. Interaction overhead could be minimized through zero-touch data collection from existing sources (vitals monitors, radio transcription), one-tap confirmation patterns, and smart defaults requiring only error correction. 

\subsubsection{Design for Infrastructure Realities and Failure Modes} We learned from participants that rural connectivity unreliability and radio quality variations require offline-first architectures where core functions (protocol support, calculations, vitals monitoring) work without connectivity, with opportunistic synchronization  (such as in \cite{miller2004consolidated}) when available. AI could provide redundancy but requires backup: augment radio with AI-enhanced MDT rather than replacing channels, provide data export for manual documentation if AI fails. Reliability thresholds should be set at levels appropriate to the stakes: extremely high accuracy (>99.9\%) for medication dosing, slightly lower thresholds for report formatting. It is important to provide graceful degradation notices (e.g. ``Vital signs trend analysis unavailable due to sensor malfunction''). 

\section{Limitations and Future Work}
This study has several limitations, which inform future work. First, our findings are based on data collected from semi-structured interviews capturing EMS clinicians' perspectives and reported experiences rather than direct observation of emergency response work. While interviews provide rich insight into EMS clinicians' concerns and decision-making processes, observational studies would complement our findings by revealing tacit practices and coordination patterns that are difficult to articulate in retrospect. Future work will include field studies observing EMS workflows to gain a more granular understanding of distributed cognition and SA maintenance mechanisms, informing more precise design requirements.

Second, as noted in Section 2.3, our findings are situated within the US EMS system. International comparative research could reveal how different staffing models (e.g., physician-staffed ambulances) and regulatory frameworks shape both the coordination challenges we identified and the design space for AI integration.
As EMS agencies begin deploying AI systems, we plan international comparative studies examining how design choices affect coordination quality, decision outcomes, and professional autonomy.

Finally, we examined perceptions of hypothetical AI systems rather than evaluating deployed technologies. As AI tools become more prevalent in EMS practice, longitudinal studies tracking actual adoption, adaptation, and impact on situational awareness and coordination will be essential. We intend to study, prototype, and evaluate AI systems implementing our five design principles, to  learn whether these approaches successfully enhance rather than degrade EMS clinicians performance under realistic conditions. Our preliminary studies will be conducted within training settings and simulations rather than in the field.

\section{Conclusions}
EMS teams maintain effectiveness through situational awareness distributed across roles, coupling pattern recognition to protocol-driven action under extreme cognitive load. AI integration threatens this tight coupling because it inserts analytical deliberation at the point where seamless recognition-to-action is essential. The five tensions EMS clinicians identified reflect structural incompatibilities between how current AI systems operate (analytical, centralized, opaque) and how emergency teams coordinate (recognition-primed, distributed, transparent).

Our five design principles address these incompatibilities by asking how AI can augment the mechanisms through which EMS teams work. We further find that this requires transparency, traceability, offline resilience, and genuine human authority. 

In summary, we find that EMS clinicians understand their needs and that treating EMS clinicians as design partners requires accepting that sometimes the most human-centered AI system is one that stays in the background.

\begin{acks}
This research was partially funded by a faculty research fellowship from the Madeleine Korbel Albright Institute for Global Affairs and by a Wellesley College Brachman Hoffman grant. We thank Navya Tiwari (Olin College), Eden Menipaz (Smith College), Zhamilya Bilyalova (Wellesley College) and Lanna Labai (University of Haifa) for supporting this work.

The authors used generative AI tools (ChatGPT, Claude, Gemini) to improve the readability of certain paragraphs, assist with table formatting, and render figures. All text, tables, and figures were written, designed, reviewed, and verified by the authors to ensure accuracy, originality, and alignment with the research findings.
\end{acks}
\bibliographystyle{ACM-Reference-Format}
\bibliography{sample-base}

\appendix

\end{document}